\documentclass[12pt]{article}
\usepackage[T2A]{fontenc}
\usepackage[utf8]{inputenc} 
\usepackage[english]{babel} 
\usepackage{amssymb,eqnarray,amsfonts,amsmath,mathtext,cite,enumerate,float} 
\usepackage{graphicx}
\usepackage{epstopdf} 
\usepackage{booktabs}
\usepackage[usenames]{color}
\usepackage{colortbl}
\title{Dynamical analysis of the Tsallis holographic dark energy models with event horizon as cut-off and interaction with matter }
\author{A. V. Astashenok and A. S. Tepliakov}
\date{}

\begin{document}

\maketitle

\begin{abstract}
The model of generalized Tsallis holographic dark energy (which is known to be particular representative of Nojiri-Odintsov HDE) with event horizon as cut-off is investigated using methods of dynamical analysis. We take into consideration possible interaction with dark energy and matter in various forms. Critical points are determined. Cosmological evolution of the Universe depends from interaction parameters. If we use event horizon scale as cutoff quasi-de Sitter expansion is possible only for interaction of type $\sim H(\alpha\rho_{de}+\beta\rho_{m})$ (where $H$ is the Hubble parameter). For interactions $\sim \rho_m  
\rho_{de} /H$ and $\sim H \rho_{m}^{\alpha}\rho_{de}^{1-\alpha}$ Universe eventually stops ($H\rightarrow 0$) or ends its existence in final singularity ($H\rightarrow\infty$). In first case fraction of dark energy tends to $1$ or constant value lesser than 1 because dynamical equilibrium between matter and dark energy is established on late times.  
\end{abstract}

\section{Introduction}

The accelerated expansion of the Universe ~\cite{1,2} is one of the main puzzles of modern theoretical physics and cosmology. The dependence of the brightness of type Ia supernovae on redshift cannot be explained by assuming that the Universe is filled with matter and radiation. It is necessary to postulate the existence of dark energy - a substance of unknown nature, which is distributed in the Universe with an extremely high degree of homogeneity and has negative pressure. As follows from observations, dark energy practically does not interact with ordinary matter in a non-gravitational way. Over the past two decades, theorists have proposed many models of dark energy. The simplest of them is the $\Lambda$CDM model, in which dark energy is nothing but a positive Einstein cosmological constant. Despite the satisfactory agreement with the data of astrophysical observations \cite{Amanullah}, \cite{Blake}, \cite{LCDM-1}, \cite{LCDM-2}, \cite{LCDM-3}, \cite{LCDM-4}, \cite{LCDM-5}, \cite{LCDM-6}, \cite{Bamba:2012cp}, \cite{LCDM-7}, this model has a number of problems at the fundamental level. The main one, the so-called problem of smallness of the cosmological constant, consists in the fact that quantum field estimates of the value of vacuum energy give a value that is 120 orders of magnitude larger than the observed one. In some interesting models of string theory, in particular, with KKLT-instantons \cite{Linde} this dramatic difference can be lowered by several tens of orders, but nevertheless it remains quite significant. It is also necessary to answer the question of why the value of the cosmological constant is comparable in order of magnitude to the observed density of matter at the present time (the coincidence problem).

Other hypotheses have also been proposed to explain the problem of dark energy. The theory of quintessence looks promising, according to which dark energy is a scalar field with an effective value of the state parameter $-1<w<-1/3$ \cite{Caldwell}, \cite{Steinhardt}, \cite{Ferramacho}, \cite{Caldwell-2}. The density of the quintessence depends on time and you can build the so-called ``tracking'' solutions for cosmological evolution that solve the coincidence problem.

A number of interesting results have been obtained in the theories of the so-called modified gravity \cite{Capozziello}, \cite{Odintsov}, \cite{Turner}, \cite{Nojiri:2010wj} in which the accelerated expansion of the Universe is caused by additional terms in the action for gravity compared to curvature.

Among hypotheses about the possible nature of dark energy, an important place is occupied by the concept of holographic dark energy (see review \cite{Wang} and references therein). The most general HDE is Nojiri-Odintsov HDE , introduced in \cite{Nojiri:2005pu}, see also \cite{Nojiri-2}. It is explicitly demonstrated in ref.31 and
\cite{Nojiri:2021jxf} that all known HDEs including Tsallis, Renyi, Barrow etc
 HDEs are just particular representatives of Nojiri-Odintsov HDE. Its theoretical basis is the holographic principle (see \cite{3}, \cite{4}, \cite{5}) and its various modifications. The holographic principle states that all physical quantities within the universe, including the density of dark energy, can be described by specifying some quantities at the boundary of the universe. Thus, only two physical quantities remain in terms of which the density of dark energy can be expressed – the Planck mass $M_{p}$ and the characteristic scale $L$. For $L$, one can take the particle horizon, the event horizon, or the reciprocal of the Hubble parameter. Using the event horizon as a scale is somewhat paradoxical. After all, with the usual approach, the dynamics of the universe is predicted based on the equations of motion and initial conditions. The use of the event horizon condition paradoxically operates based on the assumption about the nature of the future evolution of the universe. However, it seems to us that we are dealing here with a situation that is unusual rather than logically contradictory. It is appropriate to recall that even when describing the horizons and interiors of black holes, we must set the boundary conditions not in the past, but in the future \cite{Novik}. From this point of view, the approach based on the application of the principle of consistency in the future looks even less strange than the use of ''teleological'' boundary conditions in black hole physics. For classical holographic dark energy, the energy density in Planck units is $\sim L^{ -2}$ where $L$ is the characteristic scale.

A generalization of this model is proposed in \cite{Tsallis-2} (see also \cite{Tsallis}). Tsallis holographic dark energy is studied in various papers for example see \cite{Tavayef}, \cite{Jahromi}, \cite{Nojiri}. Applications of generalized holographic dark energy to modified gravity were considered in \cite{Nojiri-2}, \cite{Nojiri-3}. Previously, Tsallis model on the Randall-Sandrum brane was studied by us in \cite{AA}. In the article, the admissible values of the model parameters were determined, at which it is consistent with the data of astrophysical observations. A possible interaction between matter and holographic dark energy ($\sim H\rho_{de}$) was discussed in \cite{Qihong}. The interaction leads to the disintegration of dark energy and, at a certain intensity, to the fact that the shares of matter density and dark energy stabilize in the overall balance. The result is a quasi-de Sitter expansion of the universe instead of a possible singularity.

In this work, we analyze from the point of view of dynamical analysis a model of the Universe filled with matter, radiation, and Tsallis holographic dark energy (THDE) with event horizon as cutoff interacting with matter in various ways. Three types of interactions of the form $H (\alpha\rho_m + \beta\rho_{de})$, $\beta H \rho_{de}^{\alpha}\rho_{m}^{1-\alpha} $, $\lambda \rho_{m}\rho_{de}/H$ are considered successively. The critical points of the system of equations describing the evolution of such a Universe are found, their type is determined, and the evolution of such parameters as the Hubble parameter and the fraction of dark energy with time is studied. We compared obtained results with another model of THDE in which inverse value of Hubble parameter plays role of cutoff. 

\section{Basic equations}

Consider a spatially flat Universe with the Friedmann-Lemaitre-Robertson-Walker metric:
\begin{equation} 
\label{eq:1}
ds^2=dt^2-a^2(t)(dx^2 + dy^2 + dz^2).
\end{equation}
where $t$ is the cosmological time, $a(t)$ is the scale factor. Assume that the Universe is filled with dark energy, matter and radiation with densities $\rho_{de}$, $\rho_m$ and $\rho_{r}$ respectively. The cosmological equations for this metric can be written in the following form:
\begin{equation}
\label{eq:2}
H^{2} = \frac{1}{3}(\rho_{m}+\rho_{de}+\rho_r),
\end{equation}
\begin{equation}
\label{eq:3}
\dot{H} = -\frac{1}{2}(\rho_{m}+\rho_{de}+4\rho_r/3+p_{de}).
\end{equation}
Here the Hubble parameter is defined as $H=\dot{a}/a$. In the Tsallis model, the density of holographic dark energy is
\begin{equation}
\label{eq:4}
\rho_{de}=\frac{3C^{2}}{L^{4-2\gamma }},
\end{equation}
where $\gamma \in [1,2]$. This corresponds to particular choice of generalised cut-off in Nojiri-Odintsov HDE \cite{Nojiri:2005pu}, \cite{Nojiri-3}. The value $\gamma = 1$ corresponds to a simple model of holographic dark energy, with $\gamma=2$ we have the usual cosmological constant. As the scale $L$, consider the future event horizon:
$$L = a \int_{t}^{\infty}\frac{d{t}'}{a}.$$

If dark energy and matter interact with each other, then the continuity equations for the corresponding components take the form:
\begin{equation}
\label{eq:5}
\dot{\rho}_{m}+3H\rho_{m}=-Q,
\end{equation}
\begin{equation}
\label{eq:6}
\dot{\rho}_{de}+3H(\rho_{de}+p_{de})=Q.
\end{equation}
Here, a certain function $Q$ is introduced into the right-hand sides of the equations, which in the general case depends on time and densities. The total density $\rho=\rho_{de}+\rho_m + \rho_r$ satisfies the usual continuity equation. From the equations above, we can get an expression for the dark energy pressure $p_{de}$.
\begin{equation}
\label{eq:7}
p_{de}=-\frac{\dot{\rho}_{de}}{3H}-\rho_{de}+\frac{Q}{3H}.
\end{equation}
Then the parameter of the equation of state for dark energy is
\begin{equation}
\label{eq:8}
w_{de}=\frac{p_{de}}{\rho_{de}}=-1-\frac{\dot{\rho}_{de}}{3H\rho_{de}}-\frac{Q}{3H\rho_{de}}
\end{equation}

The holographic dark energy model with an event horizon can be consistent with observational data for various values of $\gamma$ and $C$, which makes this model quite attractive for explaining dark energy. Next, we will study the cosmological evolution of the Universe for various initial conditions, using the methods of dynamical analysis for systems of differential equations.

For $\gamma=1$ the value of $C$ is dimensionless and $L$ can be given in units of $H_{0}$, Hubble parameter at the initial moment of time. Evolution of the Universe depends solely from the values of the fractions of the densities of dark energy, radiation and matter at $t=0$ and the parameter $C$. For $\gamma\neq 1 $ the situation changes: parameter $C^{\frac{1}{\gamma-1}}$ have the same dimension as $H$. For analysis one need to know relation between this value and value of $H_{0}$. For simplicity we choose $C=1$ and therefore below we take $H$ in units of $C^{\frac{1}{1-\gamma}}$.

\section{Tsallis holographic dark energy model}

For further analysis, it is convenient to introduce the following parameters
\begin{equation}
x = \frac{\rho_m}{3H^2},\: \:\:  y = \frac{\rho_{de}}{3H^2},\: \:\:  z = \ln H, \: \: \:  \Omega_r = \frac{\rho_r}{3H^2}.
\end{equation}
The parameters $x$, $y$, $\Omega_r$ have a simple meaning of the fractions of matter density, dark energy and radiation in the overall energy balance of the Universe. The parameter $\Omega_r$ is obviously related to $x$ and $y$ by the simple relation
\begin{equation}
\Omega_r = 1 - x - y.
\end{equation}
Lets recall that the value of the Hubble parameter is measured in units of $C^{\frac{1}{\gamma-1}}$. Note a useful relation for the time derivative of the dark energy density for further calculations
\begin{equation}\label{drhodt}
\frac{\dot{\rho_{de}}}{3H^3} = (2\gamma - 4)y\left(1 - \delta\right), \quad \delta = \left(e^{2(1-\gamma)z}\frac{C^2}{y}\right)^{\frac{1}{2\gamma-4}}.
\end{equation}
Using the variable $\eta = \int H dt = \ln a$, we obtain a system of dynamic equations for $x$, $y$, $z$:
\begin{equation}
\frac{dx}{d \eta } = x - x^2 - xy(4 + (2\gamma - 4)(1 - \delta)),
\end{equation}

\begin{equation}
\frac{dy}{d \eta } = -xy + y(1-y)(4 + (2\gamma - 4)(1 - \delta)),
\end{equation}

\begin{equation}
\frac{dz}{d \eta } = -\frac{1}{2}\left(4-x-4y-(2\gamma-4)y(1-\delta)\right),
\end{equation}
When deriving the last equation, we use the obvious relation for $L/a$:
$$
\frac{d}{dt}\left(\frac{L}{a}\right) = -\frac{1}{a}.
$$
The state parameter for dark energy is calculated using the relations (\ref{eq:7}) и (\ref{drhodt}):
\begin{equation}
\omega_{de} = -\frac{1}{3}(2\gamma-4)(1-\delta) - 1.
\end{equation}
The critical points of this system of equations are found from the conditions
$$
\frac{dx}{d\eta} = 0,\quad \frac{dy}{d\eta} = 0,\quad \frac{dz}{d\eta} = 0.
$$
Obviously, there is always a critical point $x=y=0$ corresponding to the past relative to the current moment of time, when the radiation density significantly exceeded the densities of other components in the Universe. This point is the repeller. 

To search other critical points, we can explore the reduced system of equations, which is obtained from the assumption that the radiation density decreases sufficiently rapidly in the future, and therefore we can set $\Omega_r = 0$, which is equivalent to the assumption $x=1-y$. Then the first and second equations will coincide with each other.

Special analysis is required for ``infinitely distant'' (on the $y-\ln H$ diagram) points. These points include the repeller corresponding to $y=0$, $x=1$ and $H\rightarrow\infty$. Another such point - $y=0$, $x=1$ and $H=0$ - is a saddle point when there is no interaction between dark energy and matter.

To analyze the behavior of solutions near $H=0$, it is convenient to pass from the equation for $z=\ln H$ to the equation for variable $H$:
\begin{equation}
\frac{dH}{d\eta} = -\frac{H}{2}\left(4 -x - 4y -(2\gamma-4)y(1-\delta)\right).
\end{equation}
Note that for $\gamma=1$ $\delta = (y/C^2)^{\frac{1}{2}}$ and the equation for the variable $z=\ln H$ ceases to be necessary for the analysis of the cosmological dynamics of the universe. The latter will be determined solely by the value of the constant $C$ and $y_0$ - fraction of the dark energy at $t=0$.

\begin{table}[H]
\begin{center}
\begin{tabular}{@{}cccccc@{}}
\toprule
Point & x & y & $\Omega_{r}$ & Existence, behavior of $H$ \\ \midrule
1 & 0 & 0 & 1 & $H\rightarrow\infty$ (repeller) \\  \bottomrule 
2 & $0$ & $1$ & 0 & $C<1$: $H\rightarrow\infty$; $C>1$: $H\rightarrow 0$; $C=1$: $H\rightarrow\mbox{const}$\\ \bottomrule
3 & 1 & 0 & 0 & $H\rightarrow 0$ (saddle) \\  \bottomrule
4 & 1 & 0 & 0 & $H\rightarrow \infty$ (repeller) \\  \bottomrule
\end{tabular}
\caption{Critical points for the system of equations describing the evolution of the Universe filled with matter, radiation and holographic dark energy for $\gamma=1$. The first critical point is the repeller. Depending on the value of $C$, the attractor ($x=0$,$y=1$) corresponds to a different behavior of $H$. For $C=1$ the Universe expands in a quasi-de Sitter mode, $C<1$ corresponds to phantom evolution with a big rip singularity in the future, and for $C>1$ the expansion rate of the Universe tends to zero. Point 3 is a saddle point, trajectories around this point eventually lead to point 2. Trajectories in the vicinity of point 4 have a similar behavior.}
\end{center}
\end{table}
The critical points for $\gamma=1$ are listed in Table 1. The cosmological evolution essentially depends on the value of $C$: for $C>1$ the value of $z$ tends to $-\infty$, which corresponds to $H\rightarrow 0$ for $t\rightarrow\infty$. For $C<1$ we have a singularity in the future, because $z\rightarrow\infty$. Finally, $C=1$ corresponds to a quasi-desitter expansion with $H\rightarrow H_{c}$.

The case $\gamma\neq 1$ is more interesting (see Table 2). The parameter $C$ is a dimensional quantity, the critical point ($y_{0}=1$, $z_{0} = \ln C/(\gamma-1)$) is a saddle point regardless of the value of $C$. Depending on the initial conditions, the Hubble parameter $H\rightarrow 0$ at $t\rightarrow\infty$ or the Universe ends its evolution at the singularity $H\rightarrow\infty$.

\begin{table}[H]
\begin{center}
\begin{tabular}{@{}cccccc@{}}
\toprule
Point & x & y & $\Omega_{r}$ & Existence, behavior of $H$ \\ \midrule
1 & 0 & 0 & 1 & $H\rightarrow\infty$ (repeller) \\ \bottomrule 
2 & $0$ & $1$ & 0 & $H\rightarrow 0$ or $H\rightarrow \infty$ (attractor) \\ \bottomrule 
3 & $0$ & $1$ & 0 & $H=C^{\frac{1}{\gamma-1}}$ (saddle) \\\bottomrule
4 & 1 & 0 & 0 & $H\rightarrow 0$ (saddle) \\  \bottomrule
5 & 1 & 0 & 0 & $H\rightarrow \infty$ (repeller) \\  \bottomrule
\end{tabular}
\caption{Critical points for the system of equations describing the evolution of the Universe filled with matter, matter and holographic dark energy at $\gamma\neq 1$. A saddle point appears, which corresponds to the quasi-de Sitter extension for $\gamma=1$.}
\end{center}
\end{table}

\section{THDE with interaction between dark energy and matter}

If there is an interaction between matter and dark energy, the equations for the functions $x$, $y$, $z$ change as follows
\begin{equation}\label{eq1}
\frac{dx}{d \eta } = (q-x)(x-1) - xy(4 + (2\gamma - 4)(1 - \delta)),
\end{equation}

\begin{equation}\label{eq2}
\frac{dy}{d \eta } = y(q - x + (1-y)(4 + (2\gamma - 4)(1 - \delta))),
\end{equation}

\begin{equation}\label{eq3}
\frac{dz}{d \eta } = -\frac{1}{2}\left(q+4-x-4y-(2\gamma-4)y(1-\delta)\right).
\end{equation}
Here, the quantity $q$ is introduced, which is determined from the relation
$$
q = \frac{Q}{3H^3}
$$
For the dark energy state parameter, we obtain the relation
\begin{equation}
\omega_{de} = -\frac{1}{3}(2\gamma-4)(1-\delta) - 1 + \frac{q}{3y}.
\end{equation}

It is possible to analyze those critical points that appear regardless of the specific form of the function $q$ describing the interaction.

1. $H\rightarrow \infty$, $y=1$. The derivative $dy/d\eta$ is positive as $H\rightarrow\infty$ and the derivative $dz/d\eta$>0, so this point is an attractor.

2. $H\rightarrow\infty$, $y=0$. The derivative $dy/d\eta>0$ as $H\rightarrow\infty$, and the derivative $dz/d\eta$<0, so this point is a repeller.

Two another infinitely distant (on $y-\ln H$ points are ($H=0$, $y=1$) and ($H=0$, $y=0$). Write out the derivatives of the right-hand sides of the reduced system of equations. 

\begin{equation}
    \frac{df}{dy} = (1-2\gamma)(-1+2y) + y\frac{\partial q}{\partial y} + q +
\end{equation}
$$
+\delta \left((1-2y)(4-2\gamma)+1-y\right).
$$

To analyze the nature of the phase portrait around the points with $H=0$, we pass from the variable $z$ to the variable $H$. Then we get:
\begin{equation}
\frac{dH}{d\eta} = \frac{1}{H}\frac{dH}{dt} = -\frac{H}{2}\left(q + 4 -x - 4y -(2\gamma-4)y(1-\delta)\right).
\end{equation}
Let us write the derivatives of the right-hand side of the equation for $H$ and $y$ with respect to $y$ and $H$:
\begin{equation}
    \frac{df}{dH} = -2y(1-y)(1-\gamma)\frac{\delta} {H},
\end{equation}

$$
\frac{dg}{dy} = -\frac{H}{2}\left(\frac{\partial q}{\partial y}+1-2\gamma + \delta(2\gamma-5)\right),
$$

$$
\frac{dg}{dH} = -\frac{1}{2}\left(q + 3 + y -2\gamma y \right)+y\delta. 
$$
It is important that multiplication
$$
\lim_{H\rightarrow 0}\frac{dg}{dy}\frac{df}{dH} = 0,
$$
although derivative $df/dH$ can diverge. This gives us a simple equation to find eigenvalues.
$$
 \frac{df}{dy}\left|_{y=0, \delta = 0}\right. = -(1-2\gamma) + q\left|_{y=0}\right.
 $$
$$
\frac{dg}{dH}\left|_{y=0, \delta = 0}\right. = -\frac{1}{2}\left(q\left|_{y=0}\right. + 3 \right)
 $$

 $$
 \frac{df}{dy}\left|_{y=1, \delta = 0}\right. = (1-2\gamma) + \frac{\partial q}{\partial y}\left|_{y=1}\right. + q\left|_{y=1}\right.
 $$

 $$
\frac{dg}{dH}\left|_{y=1, \delta = 0}\right. = -\frac{1}{2}\left(q\left|_{y=1}\right. + 4 -2\gamma \right)
$$
Depending on the function $q$, these points can have different characters. For such interactions were $q(y=0) = q(y=1) = 0$, then the first point $y = 0, H = 0$ is a saddle point, the nature of the point $y = 1, H = 0$ is determined by derivative $\frac{\partial q}{\partial y}$.

Let us consecutively consider three types of possible interactions. The type of the $Q$ function can be derived from some general hypothetical assumptions about the possible nature of the interaction between dark energy and matter. The first and most obvious assumption is that the interaction is proportional to the density (or densities) of the components. Consider such a model below.

\textbf{Model with $Q = H(\alpha \rho_m + \beta \rho_{de})$.}

For this model, the function $q$ has the form
\begin{equation}\label{int1}
q = \alpha x + \beta y.
\end{equation}
The critical points of the system of equations and the conditions for their existence are given in Table 3 and examples of phase portraits can be seen on Fig. 1. As in the case without interaction, there is an attractor $x=0$, $y=1$, $H\rightarrow\infty$, corresponding to the singularity of the future. But for $\beta\leq 0$ and $\alpha>1-2\gamma$ another attractor appears, corresponding to the asymptotical slowing of Universe expansion. The attractor becomes a saddle point when $\beta$ changes sign and $\alpha<1-2\gamma$. The saddle point $x=1$, $y=0$, $H=0$ becomes an attractor for $\alpha<1-2\gamma$. 

\begin{table}[H]
\begin{center}
\begin{tabular}{@{}cccccc@{}}
\toprule
Point & x & y & $\Omega_{r}$ & Existence, behavior of $H$ \\ \midrule
1 & 0 & 0 & 1 & $H\rightarrow \infty$ (repeller) \\ \bottomrule
2 & $0$ & $1$ & 0 & $H\rightarrow\infty$ (attractor) \\ \bottomrule
3 & $\frac{\beta}{1-2\gamma-\alpha+\beta}$ & $\frac{1-2\gamma-\alpha}{1-2\gamma-\alpha+\beta}$ & 0 & $\beta \leq 0$, $\alpha>1-2\gamma$: $H\rightarrow 0$ (attractor) \\
& & & &  $\beta \geq 0$, $\alpha<1-2\gamma$: $H\rightarrow 0$ (saddle) \\ \bottomrule
4 & $-\frac{\beta}{3+\alpha-\beta}$ & $\frac{\alpha+3}{3+\alpha-\beta}$ & 0 & $\beta \leq 0,\alpha>-3$: $H = \left(\frac{C^2}{y_{0}}\right)^{\frac{1}{2\gamma-2}}$ (saddle) \\ \bottomrule
5 & 1 & 0 & 0 & $H\rightarrow 0$ ($-3<\alpha<1-2\gamma$ - attractor, \\ 
 &  &  &  & $\alpha>1-2\gamma$ - saddle), \\ \bottomrule 
 6 & 1 & 0 & 0 & $H\rightarrow \infty$ (repeller) \\ \bottomrule
\end{tabular}
\caption{Critical points of the system of equations (\ref{eq1})-(\ref{eq3}) for the interaction with $Q=H(\alpha\rho_m + \beta\rho_{de})$.}
\end{center}
\end{table}

A natural physical constraint should be imposed on the solution obtained for the variables $x$, $y$: these variables cannot take negative values or values greater than 1. This implies a natural interpretation of the phase trajectories ending in Fig. 1 at points with $y=1$ and different values of $H$: the realistic interaction at $y\rightarrow 1$ should asymptotically turn off and such an evolution corresponds to a quasi-de Sitter expansion of the Universe with constant $H$. It also follows, obviously, from physical considerations, to impose a requirement on the coefficient $\alpha$: its value must, in any case, exceed $-3$; otherwise, matter begins to behave like dark energy, i.e. its density will increase as the universe expands, which seems physically unreasonable. In general, from the point of view of physics, one should expect that $\alpha$ and $\beta$ should be sufficiently small. Examples of dependences of $\ln H$ and $\Omega_{de}$ from $\eta = \ln a$ are given on Figs. 2 and 3.  

\begin{figure}[H]\label{fig1}
\centering 
    \includegraphics[scale=0.18]{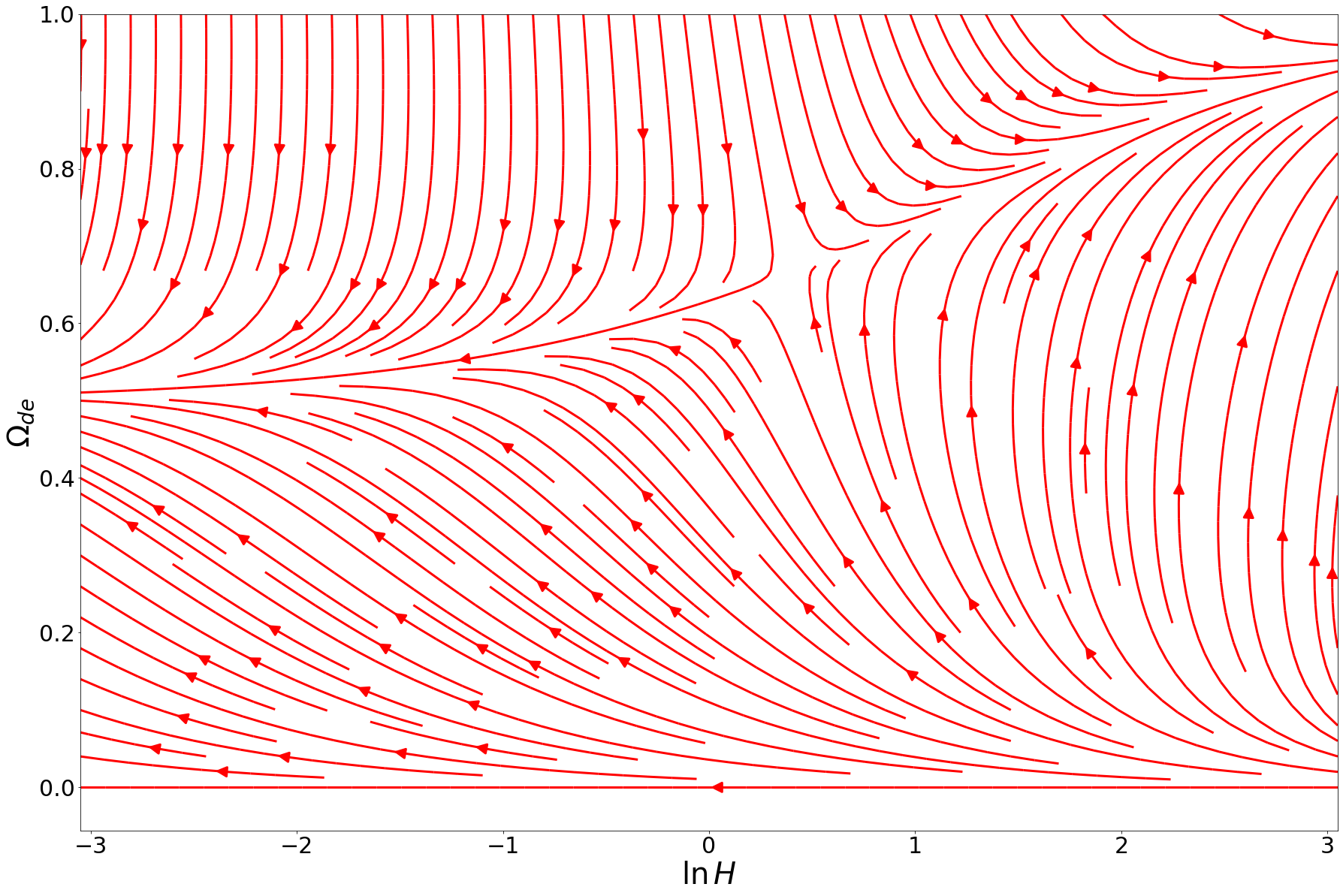}\\
    \includegraphics[scale=0.18]{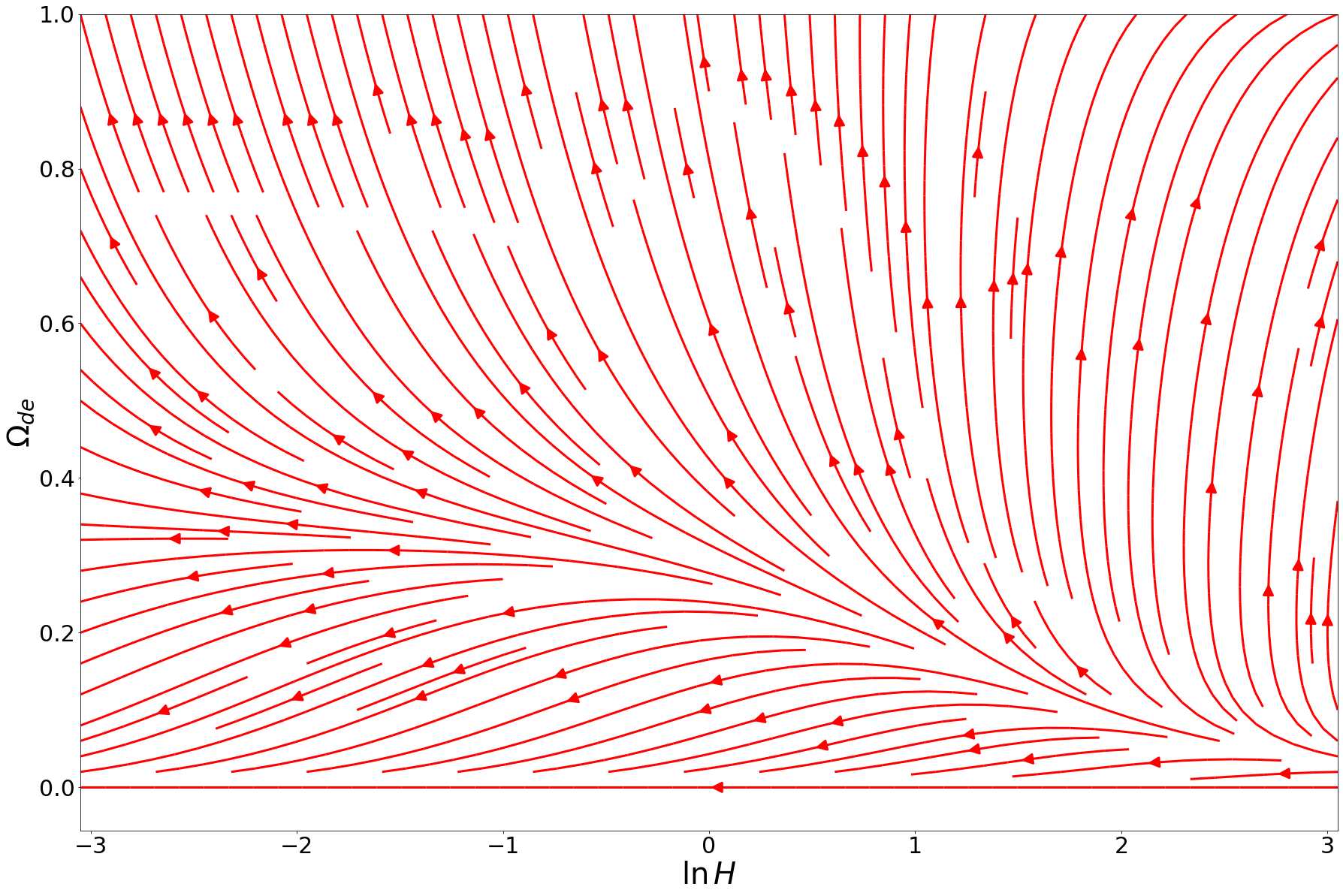}
    \caption{Examples of phase portraits for $Q = H(\alpha \rho_m + \beta \rho_{de})$. Parameters $\gamma = 1.5$ and $C=1$. a) $\alpha=-1$, $\beta=-1$: an attractor appears corresponding to a decelerating expansion with an asymptotic stop; b) $\alpha=-2.5$, $\beta=1$: the attractor corresponding to asymptotic braking becomes a saddle point, the attractor corresponding to $y=0$, $H=0$ appears.}
\end{figure}

\begin{figure}[H]\label{fig1-2}
\centering 
    \includegraphics[scale=0.21]{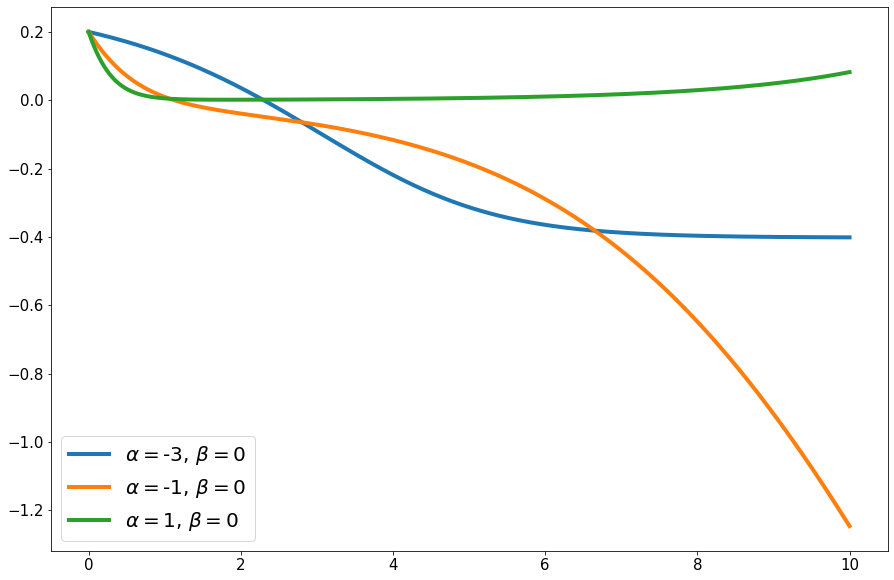} 
    \includegraphics[scale=0.21]{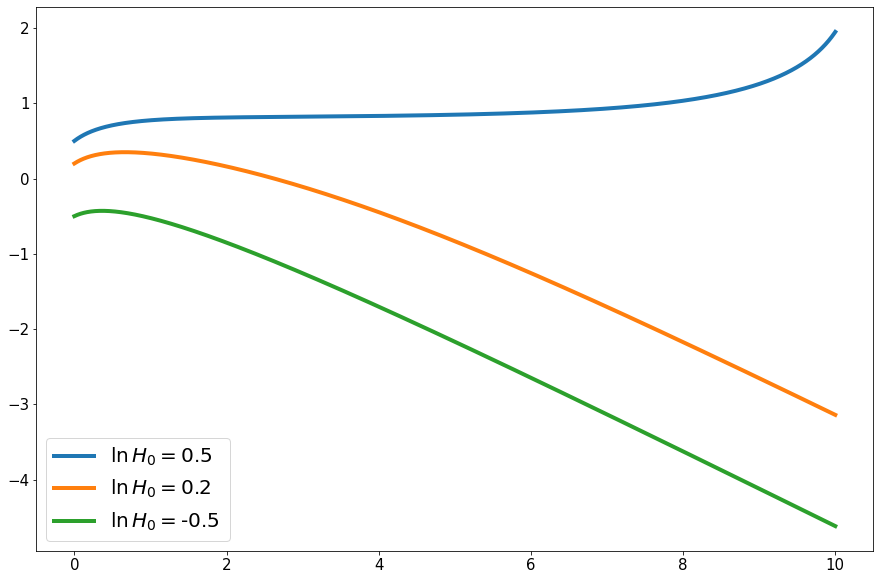} 
    \caption{Dependence of $\ln H$ from the $\eta = \ln a$: a) for different values of $\alpha$ ($\beta=0$, $\Omega_{de0}=0.7$); b) for different initial conditions and fixed values $\alpha=-1$, $\beta=-2.5$. $\Omega_{de0}=0.75$.}
\end{figure}

\begin{figure}[H]\label{fig1-3}
\centering 
    \includegraphics[scale=0.21]{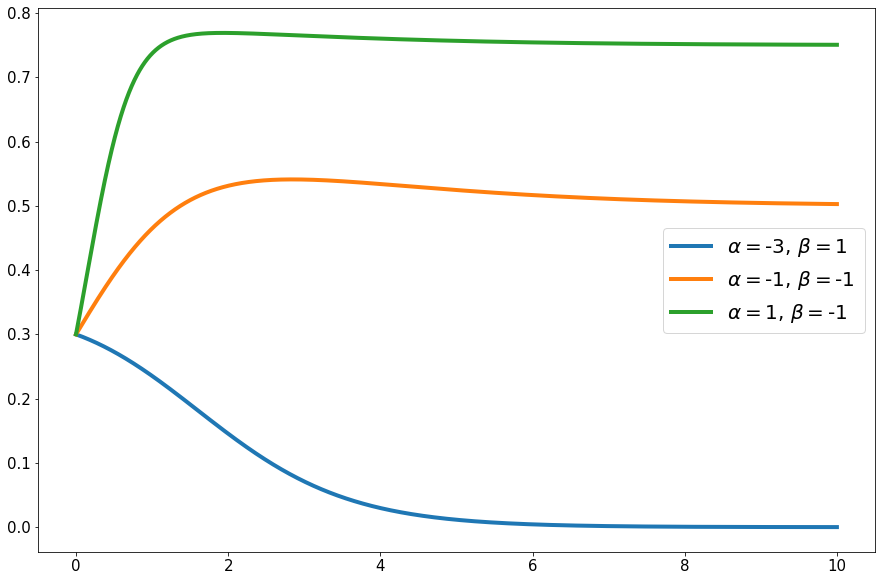}
    \includegraphics[scale=0.21]{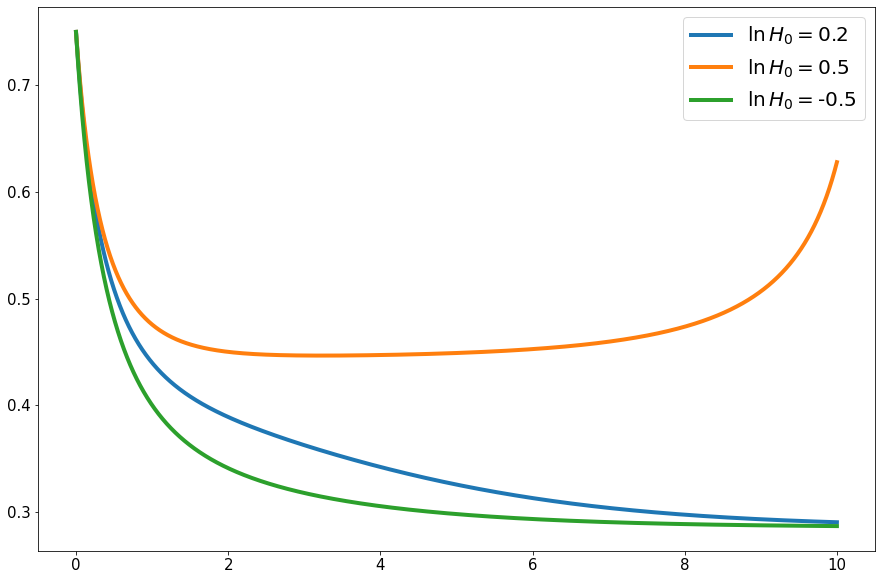} 
    \caption{Dependence of $\Omega_{de}$ from the $\eta = \ln a$ for a) different values of the parameters $\alpha$ and $\beta$ ($\Omega_{de0}=0.3$, $\ln H_{0}=0.2$; b) for different initial conditions on $\ln H_{0}$ and fixed $\alpha=-2.5$, $\beta=1$.} 
\end{figure}

\textbf{Model with $Q  = \lambda \frac{\rho_m \rho_{de}}{H}$.}

The function $q$ in this case is equal to
\begin{equation}
q = 3 \lambda x y
\end{equation}

\begin{table}[H]
\begin{center}
\begin{tabular}{@{}cccccc@{}}
\toprule
Point & x & y & $\Omega_{r}$ & Existence condition, behavior of $H$ \\ \midrule
1 & 0 & 0 & 1 & $H\rightarrow \infty$ (repeller) \\ \bottomrule
2 & 0 & 1 & 0 & $H\rightarrow \infty$ (attractor) \\ \bottomrule
3 & $0$ & $1$  & 0 & $H\rightarrow 0$ ($\lambda<(1-2\gamma)/3$ - saddle, \\ 
& & & & $\lambda>(1-2\gamma)/3$ - attractor)\\ \bottomrule
4 & $\frac{3\lambda + 2\gamma-1}{3\lambda}$ & $\frac{1-2\gamma}{3\lambda}$  & 0 & $\lambda<(1-2\gamma)/3$: $H\rightarrow 0$ (attractor) \\ \bottomrule 
5 & $\frac{\lambda+1}{\lambda}$ & $-\frac{1}{\lambda}$ & 0 & $\lambda<-1$: $H = \left({-C^2\lambda}\right)^{\frac{1}{2\gamma-2}}$ (saddle) \\ \bottomrule
6 & 0 & 1 & 0 & $H = \left({C}\right)^{\frac{1}{\gamma-1}}$ ($\lambda<-1$ - repeller, \\ 
 &  & &  & $\lambda>-1$ - saddle)  \\ \bottomrule
7 & 1 & 0 & 0 & $H\rightarrow 0$ (saddle) \\  \bottomrule
8 & 1 & 0 & 0 & $H\rightarrow \infty$ (repeller) \\  \bottomrule
\end{tabular}
\caption{Critical points for interaction with $Q  = \lambda \frac{\rho_m \rho_{de}}{H}$.}
\end{center}
\end{table}

Critical points for this interaction are listed in Table 4, and examples of phase portraits for the reduced system of equations are shown on Fig. 5. Again as in a case of first type of interaction attractor $y<1$, $H=0$ appears if $\lambda<(1-2\gamma)/3$. Point $y=1$, $H=0$ for these values of $\lambda$ becomes saddle. Also critical point $y=1$, $H=C^{\frac{1}{\gamma-1}}$ exist which for $\lambda<-1$ is unstable and if $\lambda>-1$ this point is saddle. 

\begin{figure}[H]\label{fig2}
\centering 
    \includegraphics[scale=0.12]{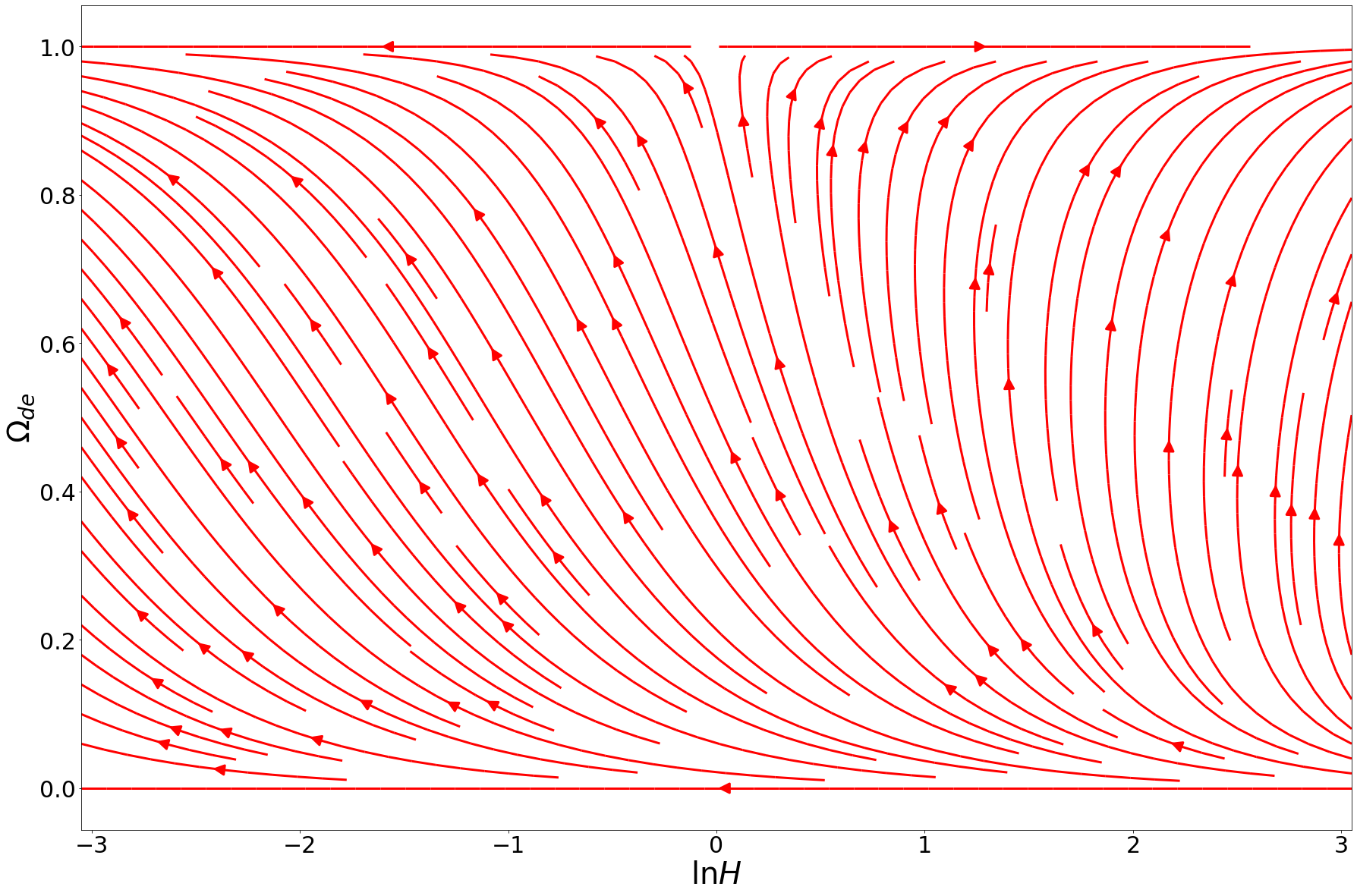}\includegraphics[scale=0.12]{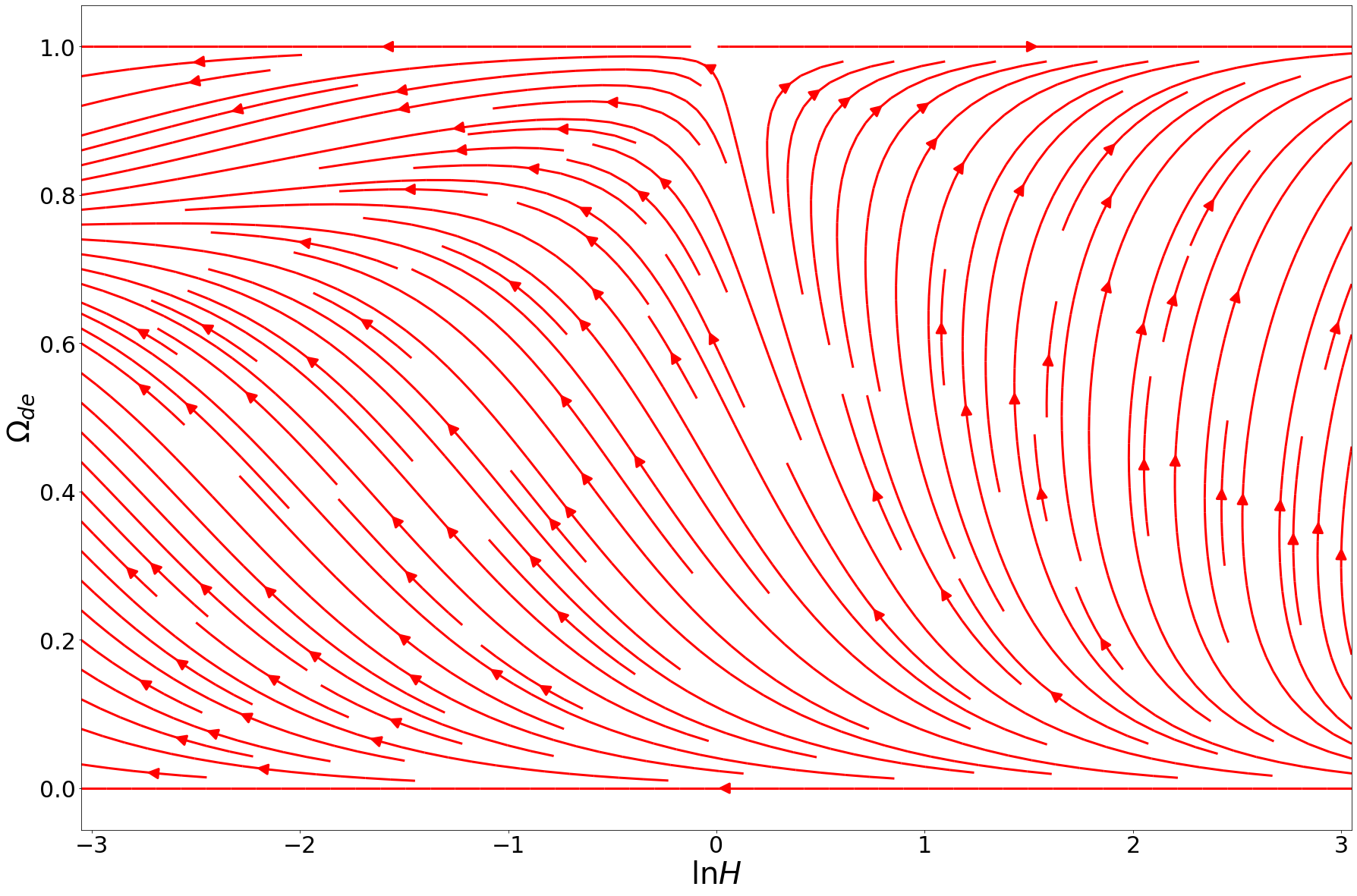}\\
    \includegraphics[scale=0.12]{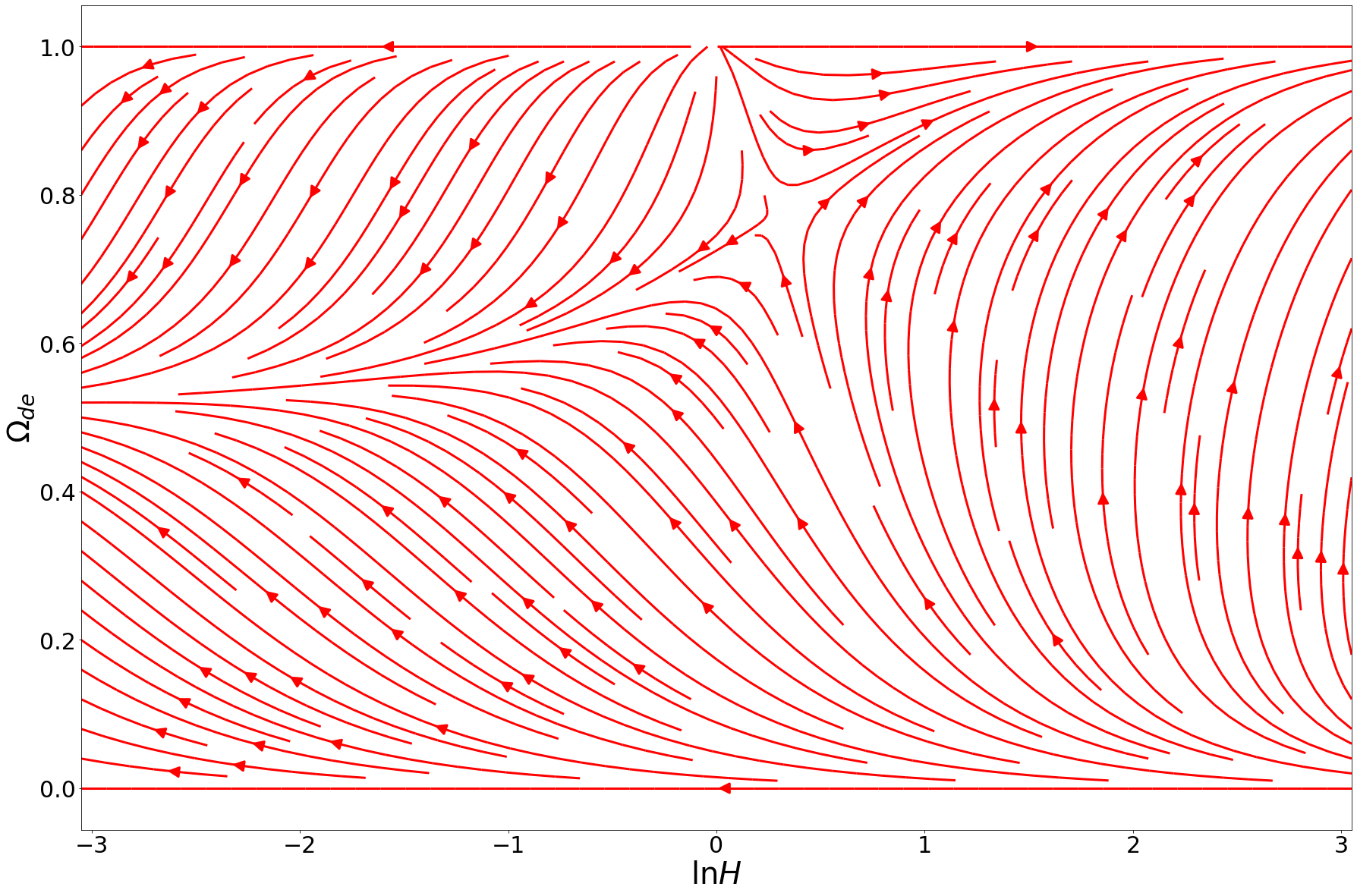} 
    \caption{Phase portraits for $Q  = \lambda \frac{\rho_m \rho_{de}}{H}$. Parameter $\gamma = 1.5$ and $C=1$. a) $\lambda=-0.5$: there are attractors 2 and 3, as well as a saddle point 6; b) $\lambda=-0.9$: point 3 becomes a saddle point, point 4 appears, which is an attractor; c) $\lambda=-1.3$: saddle point 5 appears and point 6 becomes a repeller.}
\end{figure}

\begin{figure}[H]\label{fig2-2}
\centering 
    \includegraphics[scale=0.21]{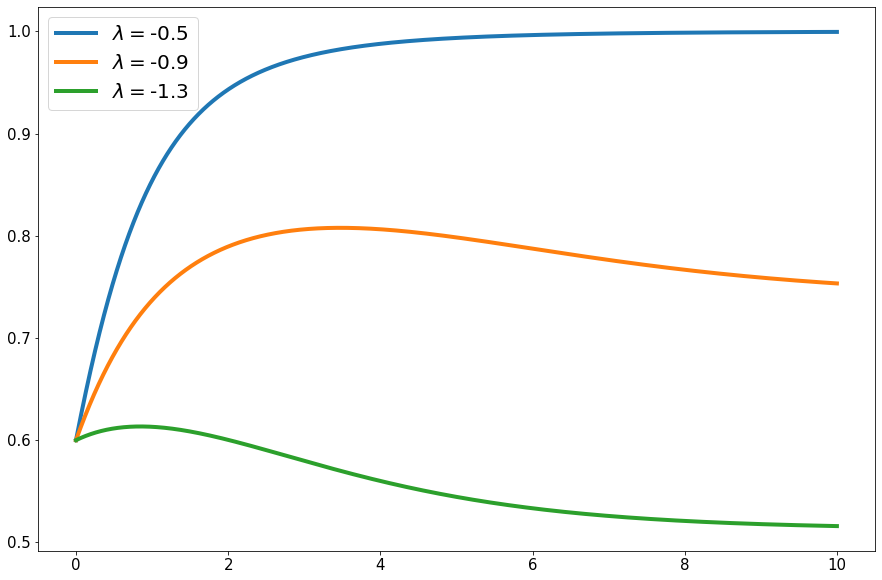}
    \includegraphics[scale=0.21]{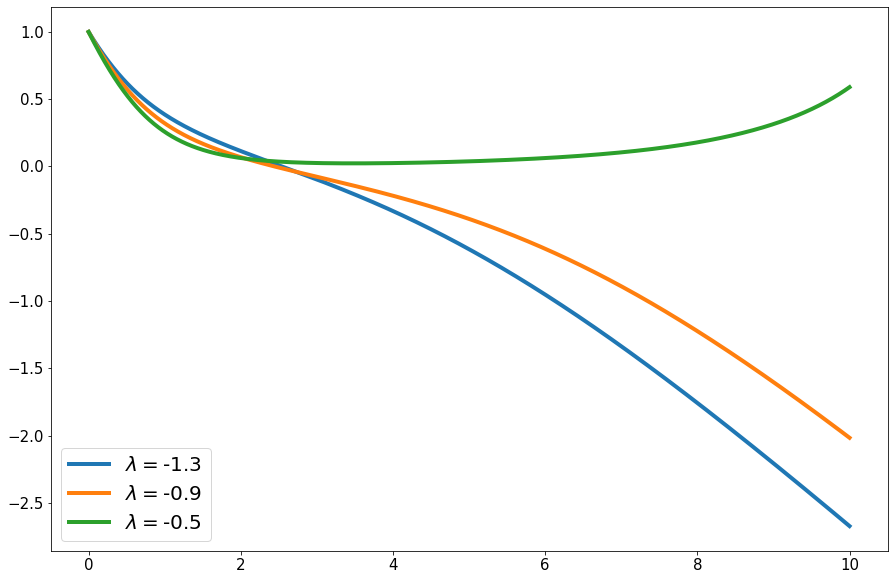} 
    \caption{Dependence of $\ln H$ from the $\eta=\ln a$ for various $\lambda$: a) for  fixed values $\ln H_0 = - 0.2$, $\Omega_{de0}=0.2$; b) for $\ln H_0 =1 $, $\Omega_{de0}=0.2$.}
\end{figure}

\begin{figure}[H]\label{fig2-3}
\centering 
    \includegraphics[scale=0.21]{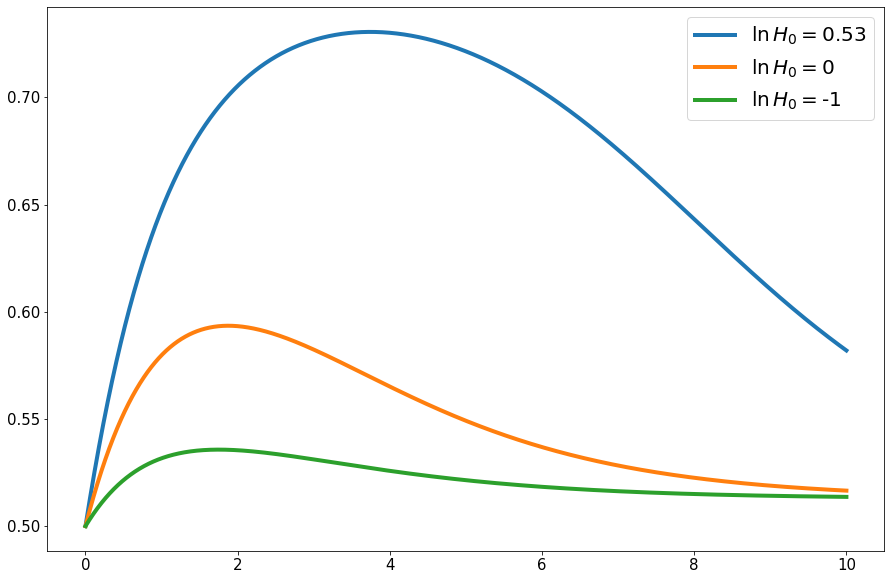} \includegraphics[scale=0.21]{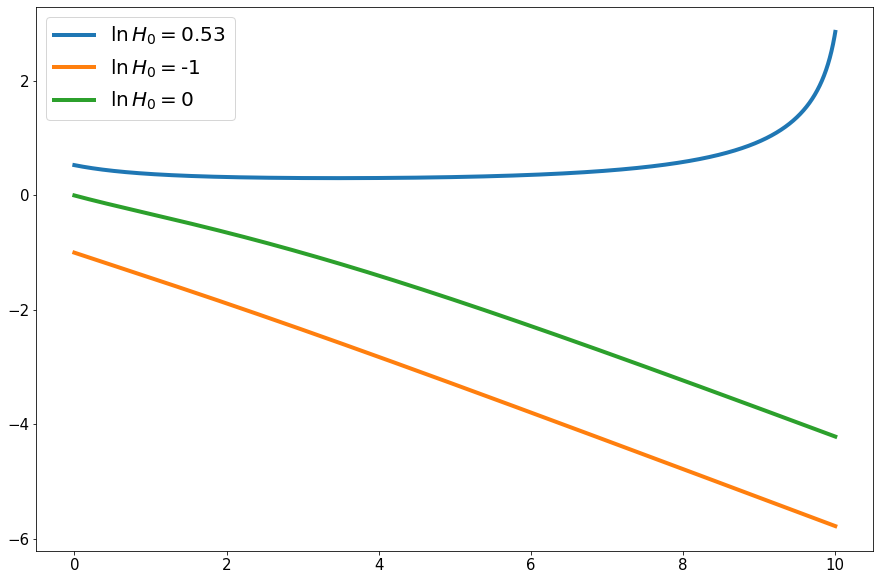} 
    \caption{Dependence of $\Omega_{de}$ from the $\eta=\ln a$ for different initial conditions on $\ln H_{0}$ and fixed values of $\lambda$ and $\Omega_{de}$: a) $\lambda = - 1.3$, $\Omega_{de0}=0.5$; b)$\lambda = - 1.3$, $\Omega_{de0}=0.52$.}
\end{figure}

\textbf{Model with $Q  = \beta H \rho_{de}^\alpha \rho_{m}^{1 - \alpha}$ ($0 <\alpha < 1$)}.

The function $q$ in this case has the form:
\begin{equation}
q = \beta x \left (  \frac{y}{x} \right )^\alpha
\end{equation}

\begin{table}[H]
\begin{center}
\begin{tabular}{@{}cccccc@{}}

\toprule
Point & x & y & $\Omega_{r}$ & Existence condition, behavior of $H$ \\ \midrule
1 & 0 & 0 & 1 & $H\rightarrow\infty$ (repeller) \\ \bottomrule
2 & 0 & 1 & 0 &  $H\rightarrow \infty $ (attractor) \\ \bottomrule
3 & 0 & 1 & 0 &  $H\rightarrow 0 $ ($\beta<0$ - saddle,  \\
 & & & & $\beta>0$ - attractor) \\ \bottomrule
4 & $\frac{(\beta/(1-2\gamma))^{1/\alpha}}{1+(\beta/(1-2\gamma))^{1/\alpha}}$ & $\frac{1}{1+(\beta/(1-2\gamma))^{1/\alpha}}$ & 0 &  $\beta<0$: $H\rightarrow 0 $ (attractor) \\ \bottomrule
5 & $\frac{1}{\left ( {-3}/{\beta} \right )^{\frac{1}{\alpha}} + 1}$ & $\frac{\left ( {-3/\beta} \right )^{\frac{1}{\alpha}}}{\left ( {-3}/{\beta} \right )^{\frac{1}{\alpha}} + 1}$ & 0 & $\beta<0$: $H=(C^2/y_0)^{\frac{1}{2\gamma-2}}$ (saddle) \\ \bottomrule
6 & 1 & 0 & 0 & $H\rightarrow 0$ (saddle) \\  \bottomrule
7 & 1 & 0 & 0 & $H\rightarrow \infty$ (repeller) \\  \bottomrule
8 & 0 & 1 & 0 & $H=(C)^{\frac{1}{\gamma-1}}$ ($\beta > 0$ - saddle, \\  
& & & & $\beta < 0$ - repeller) \\  \bottomrule

\end{tabular}
\caption{Critical points for interaction with $Q  = \beta H \rho_{de}^\alpha \rho_{m}^{1 - \alpha}$.}
\end{center}
\end{table}

For $\alpha=0$, this interaction coincides with the first one, if we put $\beta=0$ in the formula (\ref{int1}). Similarly, for $\alpha=1$ we again have a case equivalent to the previous one, provided that $\alpha=0$ in the formula (\ref{int1}). Critical points are listed in Table 5.

\begin{figure}[H]\label{fig3}
\centering 
    \includegraphics[scale=0.18]{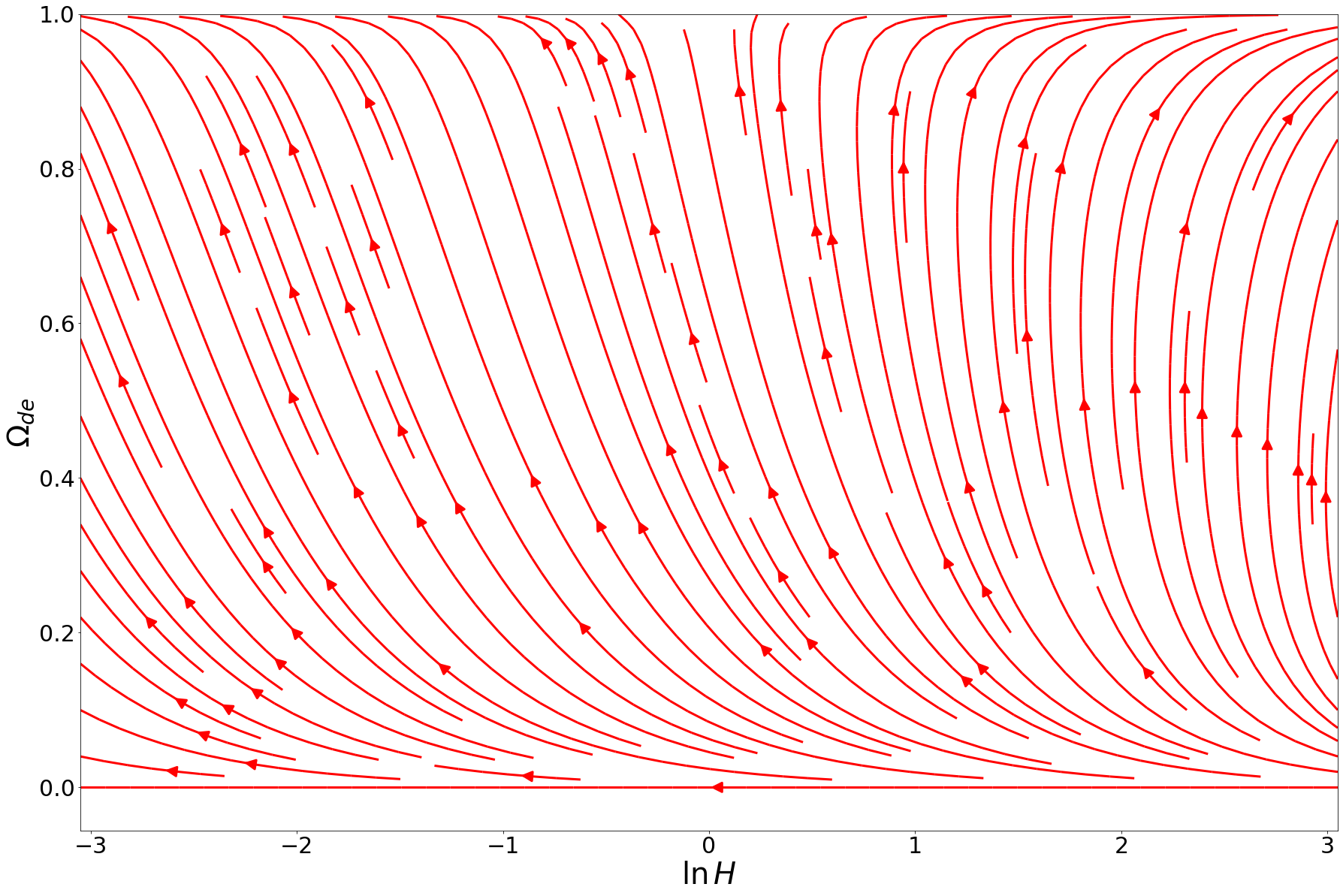}\\
    \includegraphics[scale=0.18]{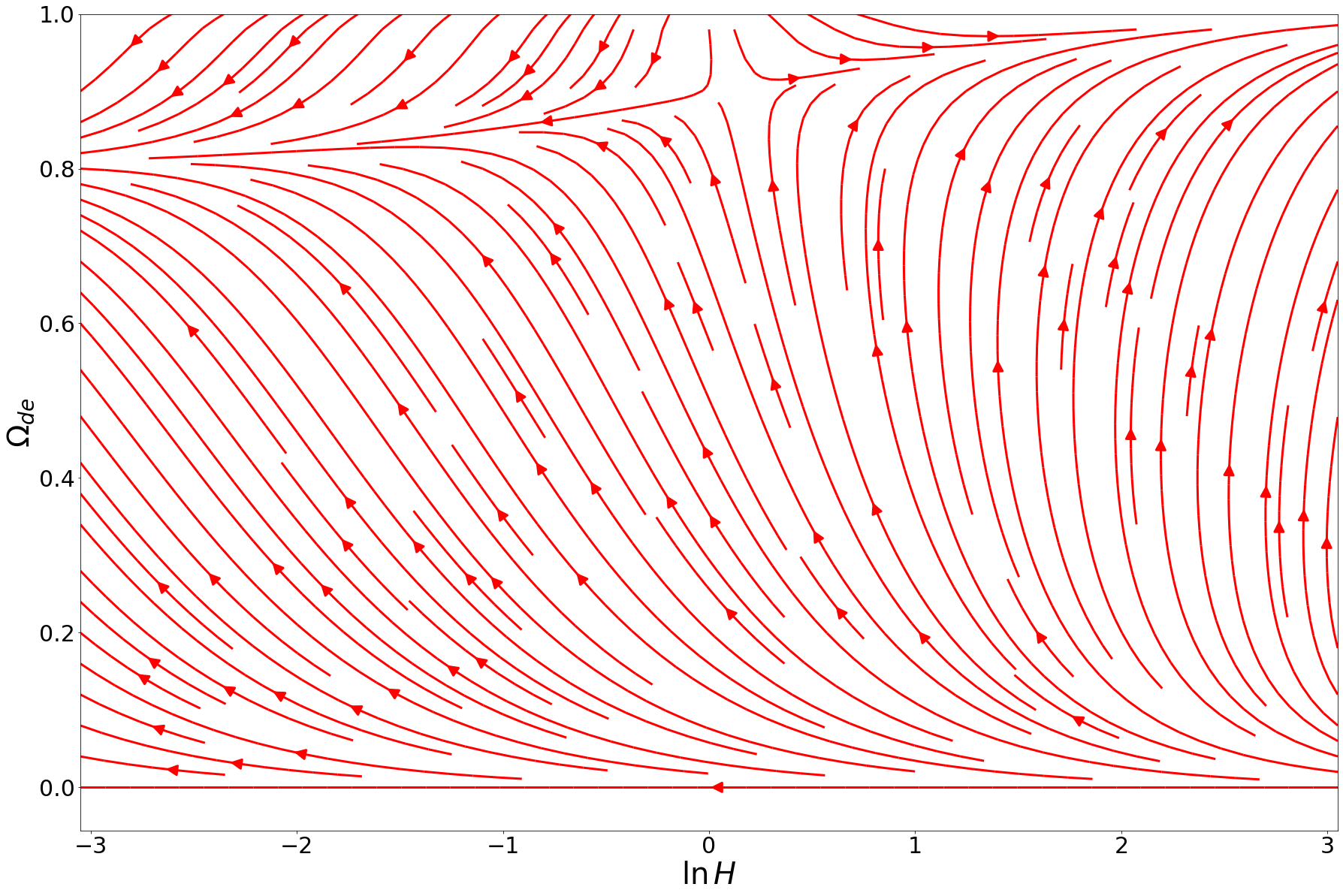} 
    \caption{Examples of phase portraits for $Q  = \beta H \rho_{de}^\alpha \rho_{m}^{1 - \alpha}$. Parameter $\gamma = 1.5$ and $C=1$. a) $\alpha=0.1$, $\beta=1$: there are critical points 2 and 3, both are attractors. b) $\alpha=0.5$, $\beta=-1$: point 3 becomes a saddle point, point 4 appears, which is an attractor, and saddle point 5 appears.}
\end{figure}

Examples of phase portraits are given on Fig. 7. As in previous cases attractor $y<1$, $H=0$ can exist. For this one need $\beta<0$. Also for $\beta<0$ saddle point with $y<1$ and $H = (C^2/y_{0})^{\frac{1}{2\gamma-2}}$ appears. For negative $\beta$ saddle point $y=1$, $H=C^{\frac{1}{\gamma-1}}$ turns into repeller.   

\begin{figure}[H]\label{fig3-2}
\centering 
    \includegraphics[scale=0.21]{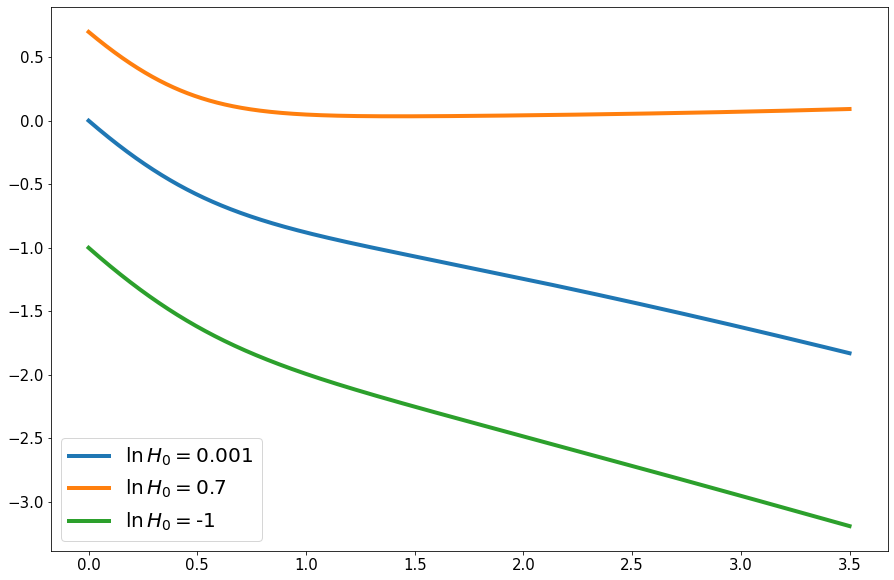}
    \includegraphics[scale=0.21]{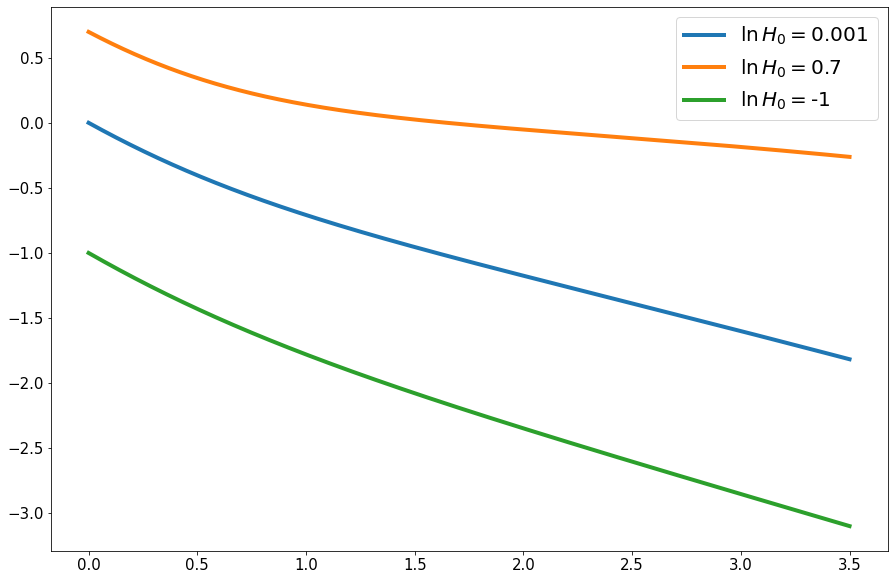} 
    \caption{Dependence of $\ln H$ from the $\eta=\ln a$ for some $\beta$ and $\alpha$: a) $\alpha=0.1$, $\beta=1$, $\Omega_{de0}=0.3$; b) $\alpha=0.5$, $\beta=-1$, $\Omega_{de0}=0.3$.}
\end{figure}

\begin{figure}[H]\label{fig3-3}
\centering 
    \includegraphics[scale=0.21]{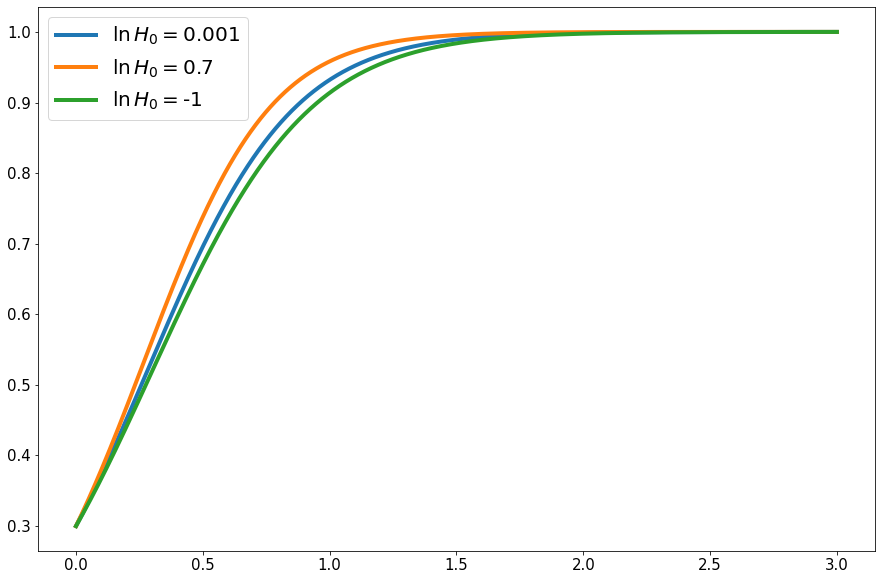} \includegraphics[scale=0.21]{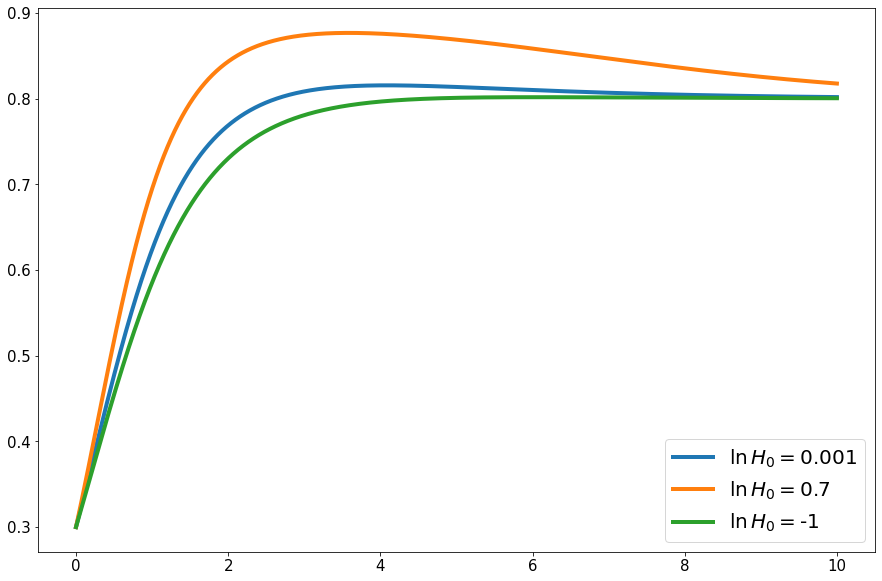} 
    \caption{Dependence of $\Omega_{de}$ from the $\eta=\ln a$ for various initial conditions on $\ln H_{0}$ and fixed another parameters: a) $\alpha=0.1$, $\beta=1$, $\Omega_{de0}=0.3 $; b) $\alpha=0.5$, $\beta=-1$, $\Omega_{de0}=0.3$.}
\end{figure}

Our consideration of various physically motivated interactions between matter and THDE shows that quasi-de Sitter expansion of the Universe for $\gamma\neq 1$ cannot be stable only for first interaction. For two other this regime is unstable. In all three cases besides attractors $y=1$, $H\rightarrow\infty$ and $y=1$, $H=0$ another attractor ($y=y_{0}<1$, $H=0$) can appear for some parameters of interacting.   

\section{Conclusion}

We investigated the cosmological dynamics of the Universe in the Tsallis model of holographic dark energy with interaction between components, using the methods of the theory of differential equations. Described below are the characteristic features of this model of dark energy compared to the usual one, in which $\rho_{de}=3C^2/L^2$. In the usual model, the future evolution of the Universe is essentially determined by the parameter $C$. At $\gamma\neq 1$ for the Universe, even without the interaction of dark energy with matter, there are two options for cosmological evolution in the future. The first corresponds to $H\rightarrow\infty$ and a singularity in the future, and the second corresponds to an asymptotic slowdown in the expansion of the Universe. The choice between the variants of evolution is actually determined by the relationship between the Hubble parameter for current time and unknown parameter $C^{\frac{1}{\gamma-1}}$.

The possible interaction between the dark components makes this picture more complex and interesting. An interaction like $Q = H(\alpha\rho_{m}+\beta\rho_{de})$ can lead to evolution, in which the ratio between matter and density remains constant, and the expansion rate tends to zero. Quasi-de Sitter expansion of the Universe at $\beta>0$ and $\alpha<1-2\gamma$ becomes possible.

The other two interactions are $Q=\lambda \rho_{m}\rho_{de}/H$ and $Q = \beta H\rho_{de}^{\alpha}\rho_{m}^{1-\alpha }$ - do not lead to the realization of the quasi-de Sitter regime in the future, but also under certain initial conditions, an asymptotic deceleration of the expansion of the Universe with a fixed ratio $\rho_m/\rho_{de}$ at $t\rightarrow\infty$ is possible.

One note that interactions with $Q=\lambda \rho_{m}\rho_{de}/H$ and $Q = H(\alpha\rho_{m}+\beta\rho_{de})$ were also considered in another model of holographic dark energy, for which $\rho_{de}=B H^{4-2\gamma}$ \cite{Qihong}. Critical points corresponding to saddle points ($H = (C^2/y_0)^{\frac{1}{2\gamma-2}}$) of the models we have considered are, with such a choice of $\rho_{de}$, stable points, which corresponds to a quasi-de Sitter expansion, but attractors appear corresponding to the asymptotic deceleration of the Universe with time. 

As a cutoff, the event horizon is taken, objections can be raised that in such a model, dark energy fluctuations grow with time. An investigation of the stability of HDE at $\gamma = 1$ was first done in \cite{Myung}, where the speed of sound in a liquid was used, which does not take into account the specifics of HDE. In \cite{LiLin} HDE is analyzed in terms of its defining property, but this requires making assumptions. \cite{LiLin} makes two assumptions: the HDE fluctuation is entirely due to the fluctuation in the size of the future event horizon (as implied by the original definition of HDE); and spherical vibration symmetry (to simplify calculations). Note that Nojiri-Odintsov holographic DE maybe generalised as holographic inflation as is demonstrated in ref. \cite{Nojiri:2019kkp}. Thus, we can conclude that HDE fluctuations are stable (there is no unlimited growth). If the quantum initial fluctuation of HDE is small enough, then it is consistent to study HDE in the late universe without fluctuations. Preliminary analysis based on the approach taken in \cite{LiLin} shows that the authors' conclusion is also valid for $\gamma \neq 1$.

This research was supported by funds provided through the Russian Federal Academic Leadership Program “Priority 2030” at the Immanuel Kant Baltic Federal University


\begin{thebibliography}{}
	
\bibitem{1} A.G. Riess, et al., Astron. J. \textbf{116} (1998) 1009.
\bibitem{2}	S. Perlmutter, et al., Astrophys. J. \textbf{517} (1999) 565.
\bibitem{Amanullah} R. Amanullah et al., Astrophys. J. \textbf{716} (2010) 712.	
\bibitem{Blake} C. Blake et al., MNRAS \textbf{418} (2011) 1707.	
\bibitem{LCDM-1} P.J.E. Peebles, B. Ratra, Rev. Mod. Phys. \textbf{75} (2003) 559.
\bibitem{LCDM-2} T. Padmanabhan, Phys. Rep. \textbf{380} (2003) 235.
\bibitem{LCDM-3} E.J. Copeland, M. Sami, S. Tsujikawa, Int. J. Mod. Phys. D \textbf{15} (2006) 1753.
\bibitem{LCDM-4} J. Frieman, M. Turner, D. Huterer, Ann. Rev. Astron. Astrophys. \textbf{46} (2008) 385.
\bibitem{LCDM-5} R. R. Caldwell, M. Kamionkowski, Ann. Rev. Nucl. Part. Sci. \textbf{59} (2009) 397.
\bibitem{LCDM-6} A. Silvestri, M. Trodden, Rept. Prog. Phys. \textbf{72} (2009) 096901.

\bibitem{Bamba:2012cp} K.~Bamba, S.~Capozziello, S.~Nojiri and S.~D.~Odintsov,
Astrophys. Space Sci. \textbf{342} (2012), 155-228.

\bibitem{LCDM-7} M. Li, X.-D. Li, S. Wang, Y. Wang, Frontiers of Physics \textbf{8} (2013) 828.
\bibitem{Linde} A. Linde, J. High Energ. Phys. 2020, \textbf{05} (2020) 76.
\bibitem{Caldwell} R.R. Caldwell, R. Dave, P.J. Steinhardt, Phys. Rev. Lett. \textbf{80} (1998) 1582.
\bibitem{Steinhardt} I. Zlatev, L. Wang, P. Steinhardt, Phys. Rev. Lett. \textbf{82} (1999) 896.
\bibitem{Ferramacho} L. Ferramacho, A. Blanchard, Y. Zolnierowsky, A. Riazuelo,  Astron. Astrophys. \textbf{514} (2010) A20.
\bibitem{Caldwell-2} R.R. Caldwell, Phys. Lett. B \textbf{545} (2002) 23.
\bibitem{Capozziello} S. Capozziello, Int. J. Mod. Phys. D \textbf{11} (2002) 483.
\bibitem{Odintsov} S. Nojiri, S.D. Odintsov, Phys. Rev. D \textbf{68} (2003) 123512. 
\bibitem{Turner} S.M. Carroll, V. Duvvuri, M. Trodden, M.S. Turner, Phys. Rev. D \textbf{70} (2004) 043528.

\bibitem{Nojiri:2010wj}
S.~Nojiri and S.~D.~Odintsov,
Phys. Rept. \textbf{505} (2011), 59-144



\bibitem{Wang} S.Wang, Y. Wang, M. Li, Phys. Rep. \textbf{696} (2017) 1.

\bibitem{Nojiri:2005pu}
S.~Nojiri and S.~D.~Odintsov,
Gen. Rel. Grav. \textbf{38} (2006), 1285-1304.

\bibitem{Nojiri-2} S. Nojiri and S.D. Odintsov, Eur. Phys. J. C \textbf{77} (2017) 528.

\bibitem{Nojiri-3} S. Nojiri, S.D. Odintsov and T. Paul, Symmetry \textbf{13} (2021) 928.

\bibitem{Nojiri:2021jxf}
S.~Nojiri, S.~D.~Odintsov and T.~Paul,
Phys. Lett. B \textbf{825} (2022), 136844.

\bibitem{3} J.D. Bekenstein, Phys. Rev. \textbf{D7} (1973) 2333.
\bibitem{4} S.W. Hawking, Commun. Math. Phys. \textbf{43} (1975) 199.
\bibitem{5} G. t Hooft, arXiv:gr-qc/9310026.
\bibitem{Novik} I.D. Novikov, V.P. Frolov, Phys. Usp. \textbf{44} (2001) 291.
\bibitem{Tsallis-2} C. Tsallis, L.J.L. Cirto, Eur. Phys. J. \textbf{C73} (2013) 2487.
\bibitem{Tsallis} C. Tsallis, J. Stat. Phys. \textbf{52} (1988) 479.
\bibitem{Tavayef} M. Tavayef et al., Phys. Lett. B \textbf{781} (2018) 195.
\bibitem{Jahromi} A.S. Jahromi et al., Phys. Lett. B \textbf{780} (2018) 21.
\bibitem{Nojiri} S. Nojiri, S.D. Odintsov and E.N. Saridakis, Eur. Phys. J. C \textbf{79} (2019) 242.


\bibitem{AA} A.V. Astashenok, A.S. Tepliakov, Int. J. Mod. Phys. D \textbf{29} (2019) 1950176.
\bibitem{Qihong} Q. Huang et al., Class. Quant. Grav. \textbf{36} (2019) 175001.



\bibitem{Myung} Y.S. Myung, Phys. Lett. B \textbf{652} (2007) 223.
\bibitem{LiLin} M. Li, C. Lin, Y. Wang, JCAP \textbf{08} (2008) 023.

\bibitem{Nojiri:2019kkp}
S.~Nojiri, S.~D.~Odintsov and E.~N.~Saridakis,
Phys. Lett. B \textbf{797} (2019), 134829.

\end{thebibliography}
\end{document}